# Mimicking Collective Firing Patterns of Hundreds of Connected Neurons using a Single-Neuron Experiment

Amir Goldental[1,†], Pinhas Sabo[1,†], Shira Sardi[1,2], Roni Vardi[2,†] and Ido Kanter[1,2,*]

[1]Department of Physics, Bar-Ilan University, Ramat-Gan 52900, Israel

[2]Gonda Interdisciplinary Brain Research Center and the Goodman Faculty of Life Sciences, Bar-Ilan University, Ramat-Gan 52900, Israel

†These authors contributed equally to this work.

*ido.kanter@biu.ac.il

**The experimental study of neural networks requires simultaneous measurements of a massive number of neurons, while monitoring properties of the connectivity, synaptic strengths and delays. Current technological barriers make such a mission unachievable. In addition, as a result of the enormous number of required measurements, the estimated network parameters would differ from the original ones. Here we present a versatile experimental technique, which enables the study of recurrent neural networks activity while being capable of dictating the network connectivity and synaptic strengths. This method is based on the observation that the response of neurons depends solely on their recent stimulations, a short-term memory. It allows a long-term scheme of stimulation and recording of a single neuron, to mimic simultaneous activity measurements of neurons in a recurrent network. Utilization of this technique demonstrates the spontaneous emergence of cooperative synchronous oscillations, in particular the coexistence of fast γ and slow δ oscillations, and opens the horizon for the experimental study of other cooperative phenomena within large-scale neural networks.**

## Introduction

One of the fundamental goals in neuroscience is to understand the mechanisms underlying the emergence of time-dependent collective activities(Silva et al., 1991;Gray, 1994;Contreras et al., 1997;Buzsaki and Draguhn, 2004;Buzsaki, 2006;Chialvo, 2010). This understanding will shed light on the way the brain reliably analyzes information and generates behavior(Klimesch, 1999;Basar et al., 2001;Wiest and Nicolelis, 2003;Kahana, 2006;Bollimunta et al., 2008;Fries, 2009;Giraud and Poeppel, 2012). The experimental accomplishment of this goal requires the following two advanced abilities. The first ability is to record from a large number of neurons over a period of seconds and minutes, which reflects the time scale of the collective network phenomena. The second ability is to know all network parameters, e.g. the network connectivity, synaptic delays and synaptic strengths (**Figure 1A**). Thus, the number of simultaneous measurements has to be in the order of the number of neurons and synapses (**Figure 1B**). Although the technology of electrophysiological measurements was significantly enhanced during the last decades, there is not yet such a technology which can record from thousands of individual neurons with a single-cell resolution(Marx, 2014), concurrently with real-time gathering of detailed network topology, including synaptic strengths and delays(Pastrana, 2012).

It is impartial to assume that the implementation of an enormous number of measurements on the network will influence its activity, and as a byproduct will modify the network parameters. Hence, as a result of many measurements, the estimated network parameters will differ from either the original or from the actual ones (**Figure 1C**). All in all, this limitation puts in question the ability to experimentally pinpoint the quantitative interplay between the network properties and its functionalities. This limitation reminds the fundamental quantum measurement difficulties(Braginsky et al., 1995), where a measurement affects the state of the system. Although here there is no physical principle that prohibits an accurate measurement, the multi-measurements are expected to modify the network and induce unavoidable learning processes, preventing flawless real-time estimations.

We present and utilize a real-time experimental long-term single-neuron stimulation and recording scheme which allows the study of the collective firing activity of a recurrent neural network, given its synaptic strengths and delays. It extends previous attempts to understand network behavior from iterative stimulation and simulation of single cells(Reyes, 2003;Lerchner et al., 2006;Brama et al., 2014;Dummer et al., 2014). Hence, the robustness of the collective firing phenomena can be examined for different sets of synaptic delays and strengths. The presented experimental scheme serves as a mirror image of the reverse engineering methods(Csete and Doyle, 2002;Gregoretti et al., 2010) where the topology of the recurrent network is estimated from its activity and verifies recent simulations and theoretical results, which predicted similar cooperative oscillations in excitatory networks(Goldental et al., 2015).

## Materials and Methods

### Experimental Procedures

**Animals.** All procedures were in accordance with the National Institutes of Health Guide for the Care and Use of Laboratory Animals and the University's Guidelines for the Use and Care of Laboratory Animals in Research and were approved and supervised by the Institutional Animal Care and Use Committee.

### *In vitro* experiments

**Culture preparation.** Cortical neurons were obtained from newborn rats (Sprague-Dawley) within 48 h after birth using mechanical and enzymatic procedures. The cortical tissue was digested enzymatically with 0.05% trypsin solution in phosphate-buffered saline (Dulbecco's PBS) free of calcium and magnesium, and supplemented with 20 mM glucose, at 37°C. Enzyme treatment was terminated using heat-inactivated horse serum, and cells were then mechanically dissociated. The neurons were plated directly onto substrate-integrated multi-electrode arrays (MEAs) and allowed to develop functionally and structurally mature networks over a time period of 2-3 weeks *in vitro*, prior to the experiments. Variability in the number of cultured days in this range had no effect on the observed results. The number of plated neurons in a typical network was in the order of 1,300,000, covering an area of about 380 mm$^2$. The preparations were bathed in minimal essential medium (MEM-Earle, Earle's Salt Base without L-Glutamine) supplemented with heat-inactivated horse serum (5%), glutamine (0.5 mM), glucose (20 mM), and gentamicin (10 g/ml), and maintained in an atmosphere of 37°C, 5% $CO_2$ and 95% air in an incubator as well as during the electrophysiological measurements.

**Synaptic blockers.** All experiments were conducted on cultured cortical neurons that were functionally isolated from their network by a pharmacological block of glutamatergic and GABAergic synapses. For each culture 20 µl of a cocktail of synaptic blockers was used, consisting of 10 µM CNQX (6-cyano-7-nitroquinoxaline-2,3-dione), 80 µM APV (amino-5-phosphonovaleric acid) and 5 µM bicuculline. This cocktail did not block the spontaneous network activity completely, but rather made it sparse. At least one hour was allowed for stabilization of the effect.

**Stimulation and recording.** An array of 60 Ti/Au/TiN extracellular electrodes, 30 µm in diameter, and spaced 500 µm from each other (Multi-Channel Systems, Reutlingen, Germany) were used. The insulation layer (silicon nitride) was pre-treated with polyethyleneimine (0.01% in 0.1 M Borate buffer solution). A commercial setup (MEA2100-2x60-headstage, MEA2100-interface board, MCS, Reutlingen, Germany) for recording and analyzing data from two 60-electrode MEAs was used, with integrated data acquisition from 120 MEA electrodes and 8 additional analog channels, integrated filter amplifier and 3-channel current or voltage stimulus generator (for each 60 electrode array). Mono-phasic square voltage pulses typically in the range of [-800, -500] mV and [60, 400] µs were applied through extracellular electrodes. Each channel was sampled at a frequency of 50k samples/s, thus the changes in the neuronal response latency were measured at a resolution of 20 µs.

**Cell selection.** A neuron was represented by a stimulation source (source electrode) and a target for the stimulation – the recording electrode (target electrode). These electrodes (source and target) were selected as the ones that evoked well-isolated, well-formed spikes and reliable response with a high signal-to-noise ratio. This examination was done with a stimulus intensity of -800 mV with a duration of 200 µs using 30 repetitions at a rate of 5 Hz, followed by 1200 repetitions at a rate of 10 Hz.

**Data analysis.** Analyses were performed in a Matlab environment (MathWorks, Natwick, MA, USA). The reported results were confirmed based on at least eight experiments each, using different sets of neurons and several tissue cultures. Action potentials were detected on-line by threshold crossing, using a detection window of typically 2-10 ms following the beginning of an electrical stimulation.

**Implementation of the mimicking scheme.** The management of the stimulation history of each of the mimicked neurons as well as the timings of their evoked spikes was done in real-time. A detailed description of the mimicking procedure is presented in the **Appendices** in the form of a short description and an algorithm. A simplified version of the scheme is presented in **Figure 6** in the form of a flowchart as well as in the **supplemental movie** (http://journal.frontiersin.org/article/10.3389/fnins.2015.00508/).

## Results

When a neuron is stimulated repeatedly, the time-lag between a stimulation and its corresponding evoked spike, the neuronal response latency (NRL), stretches gradually(Wagenaar et al., 2004;De Col et al., 2008;Vardi et al., 2014;Vardi et al., 2015) (**Figure 2A** and **Materials and Methods**). Above a critical stimulation frequency, $f_c$, which varies much among neurons(Vardi et al., 2015), this stretching terminates at the intermittent phase. This phase is characterized by large fluctuations around a constant NRL and by neuronal response failures, NRFs (**Figure 2A**). The non-zero fraction of NRFs is such that the average firing frequency is $f_c$, independent of the stimulation frequency; hence, the neuron operates similar to a low pass filter(Vardi et al., 2015) (**Figure 2B**). In addition to the preservation of the neuron's average firing frequency under periodic stimulations, the response failures were found to be statistically independent(Vardi et al., 2015) (**Figure 2C**). Specifically, for inter-stimulation-intervals that are longer than the refractory period, the firing probability is independent of the neuron's firing history. In the general stimulation scenario, aperiodic stimulations, the statistics of the NRFs were experimentally found to depend on the short-term stimulation history of the neuron, which typically consists of several stimulations only(Vardi et al., 2015) (**Figure 2D**). These effects might be an indirect result of some kind of spike-frequency adaptation(Benda and Herz, 2003) or a related mechanism.

The proposed experimental technique allows the mimicking of the activity of a neural network, given the features of the connections and the initial condition of the firing neurons. For the sake of simplicity, we first demonstrate the utilization of the proposed method using a diluted network with above-threshold synapses and with uniform delays between neurons, $\tau$. In such a case the history of a network appears as consecutive "snapshots" of the network separated by $\tau$ time-lags between them (**Figure 3A**). Each "snapshot" of the network defines which are the stimulated neurons and which neurons fire at that time. Specifically, each neuron in each snapshot belongs to one of the following three states: received a stimulation that results in an evoked spike, received a stimulation that results in a response failure, or did not receive stimulation at that time (**Figure 3B**). The neurons to be stimulated in the consecutive snapshot are determined through the network connectivity (**Figure 3B**). For example, assume neuron A is pre-synaptic to neuron B and neuron A fires at time T, consequently neuron B is stimulated at time T+$\tau$. Neurons in the network are stimulated either if their pre-synaptic neurons fired at the previous snapshot, or if they are stimulated by a stochastic noise, e.g. synaptic noise. An example is presented in **Figure 3C$_1$**, given the network dynamics until the snapshot at time T+4$\tau$, three neurons will receive a stimulation at the next snapshot, T+5$\tau$, and their response has to be examined (**Figure 3C$_1$**). The goal now is to determine whether these three neurons will fire, based on their short-term stimulation history. This task is done experimentally *using a single mimicking neuron (in vitro or in vivo)* (see **supplemental movie** and **Materials and Methods**) and is based on the following two steps:

*The mimicking step*: The current responsiveness, response susceptibility to stimulations, of a neuron from the network is mimicked by the enforcement of its short-term stimulation history on the mimicking neuron (see **supplemental movie**), e.g. three last stimulations at **Figure 3C$_{2-4}$**. After the completion of this step, the mimicking neuron will have the same responsiveness as the mimicked neuron in the current state of the network (**Figure 2D**).

*The responsive test*: $\tau$ ms after the termination of the first step, the mimicking neuron is stimulated. In case of an evoked spike, we conclude that the mimicked neuron in the network fires and this event is noted in the current snapshot.

For each stimulated neuron in the snapshot these two steps are repeated sequentially in *real-time* (**Figure 3C$_5$**), using the same mimicking neuron, until the responsiveness of all neurons in the current snapshot is determined (**Figure 3C$_6$**). After the snapshot at time T+5$\tau$ was completed, the procedure is repeated to determine the state of the next snapshot, T+6$\tau$ (**Figure 3C$_6$**), and so on.

The massive management of the stimulation history of each neuron in the network as well as their spike timings is done in *real-time* (**Materials and Methods**), demanding faster operations in at least two orders of magnitude than the time scale of $\tau$.

The realization of the proposed real-time method is first demonstrated for an excitatory network consisting of 500 neurons, using a cultured mimicking neuron (**Appendix A**). Each neuron has 2 pre- and 2 post-synaptic connections, all are above-threshold and randomly chosen, $\tau=13$ ms, with additional external stimulation Poissonian noise with a rate of 1 Hz (**Materials and Methods**). The network dynamics over ~2 seconds indicates $\delta$ oscillations of ~2.5 Hz which coexist with $\gamma$ oscillations of ~75 Hz (**Figure 4A**), which trivially originates from the resolution 1/$\tau$ (**Figure 3C**).

This prototypical real-time technique is realized in a more realistic biological network, consisting of sub-threshold synapses as well (**Appendix B** and **Figure 4B**). The excitatory network consists of N=500 neurons, where each neuron has *50 pre- and 50 post-synaptic connections* with $\tau=15$ ms, with an additional 1 Hz Poissonian noise (stimulations). An above-threshold stimulation requires the firing of at least 4 pre-synaptic neurons, or a stimulation originated from the noise. Results indicate $\delta$ oscillations of ~0.8 Hz which coexist with $\gamma$ oscillations of ~65 Hz (**Figure 4B**), which again originates from the resolution 1/$\tau$.

The generalization of the proposed real-time technique for networks with a continuous distribution of connection delays requires a complicated procedure, since the scheme of discrete time snapshots (**Figure 3B**) is not valid in this case. The advanced procedure requires the management of the stimulation history and the timings of the evoked spikes of all neurons in a continuous time manner. The mimicking process per neuron is similar, however, technically the complexity of the algorithm is enhanced since the constraint of specific simulations and firing times is released and occur in continuous time. Utilization of the continuous scheme indicates the coexistence of δ and spontaneous γ oscillations (**Appendix C** with **Figure 4C** and **Appendix D** with **Figure 5**), where the periods of collective firing are slightly smeared as a result of continuous connection delays.

In the case where all connection delays are equal to τ, the GCD of loops of such random networks is expected to be equal to τ(Kanter et al., 2011;Vardi et al., 2012a;Vardi et al., 2012b). In such a case, neurons will fire in synchrony every τ, therefore forming γ oscillations with frequency of 1/τ (**Figure 4A,B**). On the other hand, in case of random continuous connection delays, the GCD vanishes and no synchrony is expected beyond the δ oscillations. Our results clearly indicate that the nontrivial high frequency synchrony is dominated by the average delay, i.e. the *spontaneously originated γ oscillations* have the frequency of 1/(average delay) as also observed in simulations(Goldental et al., 2015). The distribution of the connection delays affects only the quality of the synchrony (**Figure 4C**).

Mimicking the dynamical behavior of a network consisting, for instance, of thousands of neurons over several seconds requires real-time stimulations and recordings of the mimicking neurons over several hours. Specifically, the real-time duration of the experiment is equal to the number of stimulations occurred dynamically in the network, multiplied by the time it takes to mimic a neuron. For illustration, in **Figure 4A**, a network of N=500 neurons is mimicked for 2 seconds. Since $f_c$=5 Hz each one of the neurons in the network was stimulated approximately (2 seconds)·(2$f_c$)=20. To mimic once a stimulated neuron in the network, requires approximately 0.4 seconds. Hence, the total real-time of the experiment with a single mimicking neuron is expected to be 20·N·0.4 seconds = 4000 seconds, which is indeed close to ~3700 seconds (**Figure 4D**). During this period, the mimicking neuron remains in the intermittent phase as indicated by the large fluctuations of the NRL and the response failures which resulted from the high stimulation rate (**Figure 4D**).

## Discussion

The presented experimental results verify recent simulations and theoretical work which predicted such oscillations(Goldental et al., 2015). The experimental scheme presents more reliable evidence since it takes into account biological time dependent fluctuations in the responsiveness of neurons and variations in the neuronal critical frequency, as opposed to the simulations and theory.

Currently, there are some limitations to the proposed mimicking method, which is based on short-term neuronal dynamics. Long-term effects and synaptic plasticity are ignored, however they are not expected to dominate the dynamics of the network within several seconds (**Figures 4 and 5**). It might be possible to introduce synaptic dynamics, excitatory and inhibitory, to the mimicking process by stimulating and recording from coupled neurons through synaptic connections, using patch clamp technique(Debanne et al., 2008). Currently this kind of dynamics is simplified to excitatory electrical pulses.

Experimental difficulties arise when the mimicked network is composed of thousands of neurons. Primarily, the experimental time scales linearly with the size of the network, hence it is expected to exceed several hours. Preliminary results (not shown) indicate that it is possible to mimic a network of one or two thousands neurons, however sailing towards much larger systems is in question. A possible bypass to this obstacle, and a way to mimic more heterogeneous networks, with several types of neurons, is to implement several mimicking neurons in parallel, however it will require the realization of a much more complicated experimental scheme.

The idea of mimicking network dynamics using a single neuron was previously demonstrated for feed-forward networks(Reyes, 2003), where the parameters of the network are adjusted to control the activity and the mimicking process does not take into consideration short-term neuronal plasticity. This work, on the other hand, examines recurrent random networks, where the parameters are independent of the stability of the firing rates. In addition, the average delay between successive layers in a feed-forward network is irrelevant for the dynamics, since it only shifts the time of the activity by a constant. In contrast, in recurrent networks, the exact delay times are important since each neuron is affected by many delay loops. Hence, the implementation of the mimicking process of a recurrent network consisting of short delays of several milliseconds is a challenge. Additionally, as a result of the many loops each neuron is revisited many times through the dynamics. Hence the mimicking process is done many times per neuron and keeping the network parameters fixed is essential to describe the dynamical properties of the recurrent network.

The proposed real-time experimental method can also be used to mimic the firing patterns of large recurrent neural networks *in vivo*, based on long-term scheme of stimulation and recording of a single neuron *in vivo*(Brama et al., 2014). Nevertheless, the real-time management of the *in vivo* mimicking process, where delays are several milliseconds only, is still an experimental challenge.

The presented experimental technique to use a long-term experiment on a single node in order to mimic the parallel activity of a large scale network is applicable to a variety of networks with propagation delays, where the nodes exhibit a finite or no memory of the preceding conditions. Thus, this technique is expected to be relevant to a wide range of networks that play a key role in other fields such as physics, biology and economics.

## Acknowledgments

This research was supported by the Ministry of Science and Technology, Israel. The authors declare no competing financial interests.

## Author contribution

A. G. developed the theoretical framework and compared experimental results to simulations. P. S. developed and designed the interface for the real-time experiments. R. V. and S. S. prepared the tissue cultures and the experimental materials. All authors performed the experiments and analysed the data. I. K. supervised all aspects of the work. All authors discussed the results and commented on the manuscript.

## Appendix A

**Experimental scheme: Discrete time, above-threshold connections (e.g. Figure 4A):**

A mimicked network consists of N neurons. Each neuron has K randomly selected pre- and post- above-threshold synaptic connections, and all of their delays are $\tau$. Additional external noise with a rate of $F_{noise}$ was given. Typical values are K=[2, 10], $F_{noise}$=[0.5, 1.5] Hz, and $\tau$=[10, 20] ms. The stimulation sequence for the mimicking neuron scheme consists of the last M stimulations given to the neuron. M is determined by the neuron's short-term memory, where typically M=3.

The initial conditions of stimulation times were generated for each neuron independently, by randomly choosing M+1 sorted integers in the range [2-round((M+1)/(2$\tau$·$f_c$)),1], where $f_c$ is the critical frequency of the neuron. The random numbers represent the stimulation times in the history of the neuron preceding the experiment, in the form of multiples of $\tau$. Accordingly, the scheme starts at time $\tau$ and the first neurons that are stimulated are those whose maximal integer equals 1.

In **Figure 4A**, the values were N=500, K=2, $F_{noise}$=1 Hz, M=3 and $\tau$=13 ms. The mimicking neuron had $f_c$=5 Hz, and was given 1500 pre-scheme stimulations with a rate of 2$f_c$ to reach the intermittent phase.

The mimicking scheme is done using the following procedure (a simplified flowchart is presented in **Figure 6**):

Parameters:

**N** - Number of neurons.

**K** - Number of pre- and post-synaptic sub-threshold connections.

**M** - The number of stimulations in the stimulation sequence (the neuron's "memory").

**$F_{noise}$** - Noise frequency.

**$\tau$** - The delay time between connected neurons.

**$f_c$** - The critical frequency of the mimicking neuron.

**Edges** - A **N×K** connectivity matrix.

Variables:

**Queue1**, **Queue2** - An empty 1×**N** array.

**StimulationsData** - A **N**×(**M**+1) matrix that holds the stimulation times.

**Counter1**, **Counter2** - A 1×**N** empty array.

**SnapshotNo.** - An integer denoting the currently mimicked snapshot.

1. Load connectivity to **Edges** and initial conditions to **StimulationsData**.
   **StimulationsData**(**n**,**m**) indicates a stimulation to neuron **n** at snapshot number
   **StimulationsData**(**n**,**m**).
   Assign **SnapshotNo.**=1.
   Insert to **Queue1** all the neurons that are stimulated at **SnapshotNo.**=1, according to the
   initial conditions (the maximal random integer is 1, i.e. the stimulation time equals $\tau$).

2. For each neuron **n** in **Queue1**:
   - The mimicking neuron is stimulated **M** times, with the **M** corresponding inter-
   stimulation intervals from **StimulationsData**(**n**).
   - Another stimulation is given after the appropriate inter-stimulation-interval. This
   stimulation will be also the first stimulation for the next mimicking process.
   - If the stimulation results in an evoked spike:
     Add the row **Edges**(**n**) to **Queue2**, without repetitions.
   - Replace the minimal value of **StimulationsData**(**n**) with **SnapshotNo.**.

3. Generate noise in **Queue2**: Add random neurons to **Queue2** with a probability of $F_{noise}\tau/N$ for
   each neuron.

4. Clear **Queue1**, move **Queue2** into **Queue1** and clear **Queue2**.

5. Increment **Snapshot No.** by 1 and Go to clause 2.

## Appendix B

**Experimental scheme: Discrete time, sub-threshold connections (e.g. Figure 4B):**

This scheme is similar to the scheme presented in **Appendix A**, but also incorporates sub-threshold
synaptic connections.

All synaptic connections are sub-threshold, and typically the number of postsynaptic connections per neuron, K, is between 10 and 50. $K_{min}$ denotes the minimal number of simultaneous sub-threshold stimulations that can result in an evoked spike, typically between 2 and 5. During the procedure if the number of pre-synaptic neurons that fired in the previous snapshot is above $K_{min}$, the stimulation is above-threshold.

Above-threshold stimulation has a strength of -800 mV, with a duration of 200 μs. The strength of a sub-threshold stimulation is in the range of [-200 -500] mV with a duration of 200 μs. The memory of the stimulations accounts also for sub-threshold stimulations, where typically M=8. Initial conditions of stimulations are generated randomly similar to the first scheme, with additional below-threshold stimulations.

In **Figure 4B**, the values are N=500, K=50, $K_{min}$=4, $F_{noise}$=1 Hz, M=8, and τ=15 ms. The strength of the sub-threshold stimulation was -300 mV. The mimicking neuron had $f_c$=2.1 Hz, and was given 1000 pre-scheme stimulations with a rate of $2f_c$ to reach the intermittent phase.

The mimicking scheme is done using the following procedure:

Parameters:

    **N** - Number of neurons.

    **K** - Number of pre- and post-synaptic sub-threshold connections.

    **M** - The number of stimulations in the stimulation sequence (the neuron's "memory").

    **$K_{min}$** - The number of minimal concurrent stimulations that would result in an above threshold excitatory response.

    **$F_{noise}$** - Noise frequency.

    **τ** - The delay time between connected neurons.

    **$f_c$** - The critical frequency of the mimicking neuron.

    **Edges** - A **N**×**K** connectivity matrix.

Variables:

**Queue1**, **Queue2** - An empty 1×**N** array.

**StimulationsData** - A **N**×(**M**+1)×2 matrix that holds the stimulation times and strengths.

**Counter1**, **Counter2** - A 1×**N** empty array.

**SnapshotNo.** - An integer denoting the currently mimicked snapshot.

1. As in **Appendix A**.

2. For each neuron **n** in **Queue1**:

    If **Counter1**(**n**)>=**K**$_{min}$, the stimulation is above-threshold:

    - The mimicking neuron is stimulated **M** times, with the **M** corresponding inter-stimulation-intervals.

    - Above-threshold stimulation is given after the appropriate inter-stimulation-interval. This stimulation will be also the first stimulation for the next mimicking process.

    - If the stimulation results in an evoked spike:

       For each neuron **m** from the row **Edges**(**n**), add **m** to **Queue2**, if m already exists in **Queue2**, add 1 to **Counter2**(**m**).

    Replace the stimulation with the minimal time value in **StimulationsData**(**n**) with the time value of **SnapshotNo.** and the appropriate strength.

3. Generate noise in **Queue2**: Add random neurons to **Queue2** with a probability of **F**$_{noise}$τ/N for each neuron, and add **K**$_{min}$ to **Counter2** accordingly.

4. Clear **Queue1**, move **Queue2** into **Queue1** and Clear **Queue2**.

    Clear **Counter1**, move **Counter2** into **Counter1** and Clear **Counter2**.

5. Increment **SnapshotNo.** by 1 and Go to clause 2.

# Appendix C

**Experimental scheme: Continues time, above-threshold connections (e.g. Figure 4C):**

A continuous time version **Appendix A**. The continuity of this scheme is limited by the machine cycle, 20 µs in our implementation. All synaptic connections are above-threshold, where typically K=2 and M=3.

For each connection, $\tau$ was chosen randomly from [$\tau_{min}$, $\tau_{max}$], where typically $\tau_{min}$ is between 8 and 12 ms and $\tau_{max}$ is between 12 and 20 ms. Initial conditions are constructed by choosing M random delays using exponential distribution with a rate of 2**$f_c$**.

In **Figure 4C**, the values are N=500, K=2, F$_{noise}$=0.5 Hz, M=3, $\tau_{min}$=8 and $\tau_{max}$=12 ms. The mimicking neuron had $f_c$=3 Hz, and was given 700 pre-scheme stimulations with rate of 2$f_c$ to reach the intermittent phase.

The mimicking scheme is done using the following procedure:

Parameters:

**N** - Number of neurons.

**K** - The number of pre- and post-synaptic above-threshold connections.

**M** - The number of stimulations in the stimulation sequence (the neuron's "memory").

**F$_{noise}$** - Noise frequency.

**$f_c$** - The critical frequency of the mimicking neuron.

**$\tau_{min}$, $\tau_{max}$** - The time range of the delays.

**Edges** - A **N×K** matrix containing information of nodal connections and their delays. The delays are randomly chosen in the range [$\tau_{min}$, $\tau_{max}$].

Variables:

**StimulationsData** - A **N×30** matrix that holds the stimulation times.

**T** - Mimicked time.

**Counter** - A 1×**N** array of zeros.

1.  Load connectivity to **Edges** and initial conditions to **StimulationsData**.
    Assign **T**=0.

2.  Find a neuron **n** such that **StimulationsData**(**n**,**Counter**(**n**)) is minimal but also greater than **T**. If no neuron is found - go to clause 4.
    Assign **T**=**StimulationsData**(**n**,**Counter**(**n**)).

3. The mimicking neuron is stimulated **M** times, with the **M** corresponding inter-stimulation-intervals of **n** (taken from **StimulationsData**).

Another stimulation is given. This stimulation will be also the first stimulation for the next mimicking process.

If the stimulations results in an evoked spike:

> For each neuron **j** from the row **Edges**(**n**), add the sum of the spike's time and neuron **j**'s delay to **StimulationsData**(**j**,**Counter**(**j**)).

Increment **Counter**(**n**) by 1, assign **Counter**(**n**)=**Counter**(**n**) mod **N**.

4. Generate noise in **StimulationsData**: A stimulation every 0.1 ms with a probability of $10^{-4}*F_{noise}$ to each neuron.

5. Go to clause 2.

## Appendix D

**Experimental scheme: Continuous time, Sub-Threshold connections (e.g. Figure 5):**

A continuous time version **Appendix B**. In the end of each mimicking process, if the last stimulation results in an evoked spike, a weak stimulations will be given to each one of the post synaptic neurons of the mimicked neuron. If two weak stimulations arrive with a short lime-lag between them, they will be merged to one strong stimulation. The procedure could be generalized to account for stricter temporal summations, where more than two weak stimulations are required to fall within a short time window in order to yield a spike ($K_{min}$ can be defined similarly to **Appendix B**) - the current implementation requires only two stimulations. The program advances in segments of time – **SegmentLength**, such that neurons whose current segment does not contain any above-threshold stimulations will not be mimicked at the current step. A stimulation sequence consists of the short term stimulation history of a neuron, [T - **LookBack**, T], where T is the time of the current time mimicked and **LookBack** is a pre-determined time constant. There is an essential difference between this procedure and all the other procedures that should be stressed out. In the previous procedures the times of the last M stimulations were taken into account, disregarding the length of the inter-stimulation-intervals. This procedure, on the other hand, takes into account all and only the stimulations that occurred within the last **LookBack** seconds.

Initial conditions are random times in the range [-LookBack-$\tau_{max}$, 0], corresponding to the stimulations times. Only a third of the stimulations are above threshold stimulations. The average stimulation rate is $3f_c$.

In **Figure 5**, the values are N=350, K=13, NoiseNum=3, SegmentLength=4.5 ms, LookBack=3.5 s, MinGap=5.5 ms, $\tau_{min}$=6 ms and $\tau_{max}$=10 ms. The mimicking neuron had $f_c$=4.2 Hz, and was given pre-scheme stimulations 700 stimulations with rate of $2f_c$ to reach the intermittent phase.

The mimicking scheme is done using the following procedure:

Parameters:

**N** - Number of neurons.

**K** - The number of pre- and post-synaptic sub-threshold connections.

**$f_c$** - The critical frequency of the mimicking neuron.

**$\tau_{min}$, $\tau_{max}$** - The time range of the delays.

**SegmentLength** - The time segment in which the program advances, has to be smaller than the minimal neuronal response latency.

**NoiseNum** - The number of random neurons stimulated at every segment.

**LookBack** - The length of the mimicking sequence.

**MinGap** - The maximal time-lag between two weak stimulations which results in one merged supra-threshold stimulation.

**Edges** - **N×K** matrix, containing information of nodal connections and their delays. The delays are randomly chosen in the range [$\tau_{min}$, $\tau_{max}$].

Variables:

**StimulationsData** - a data structure that holds the times of weak and strong stimulations.

**IsStimulated** - **N×m** matrix, the first index indicates a neuron in the network and the second index indicates a Segment Index. **IsStimulated**(i,**SegmentIndex**)=1 means that the neuron is stimulated above-threshold at the segment Index **SegmentIndex**, otherwise it is not.

**Next** - a FIFO with the neurons that should be stimulated at the current segment.

**SegmentIndex** - An index indicating the current segment (the time step being mimicked).

1. Load connectivity to **Edges** and initial condition to **StimulationsData** and **IsStimulated**.

   Assign **SegmentIndex=**-1.

2. **SegmentIndex=SegmentIndex**+1.

   Add to **Next** all neurons **i**, such that **IsStimulated** (**i**,**SegmentIndex**)=1.

   Generate random above-threshold stimulations according to **NoiseNum**.

3. For each neuron **i** in **Next**:

   - Wait several milliseconds.

   - Stimulate the mimicking neuron with all the stimulations of **i** in the time range [(**SegmentIndex*SegmentLength**-**LookBack**), **SegmentIndex*SegmentLength**].

   - If a spike occurred after the mimicking stimulation sequence:

      - Update **StimulationsData** for the post synaptic neurons of the mimicked neuron to receive a weak stimulation at time **SegmentIndex*SegmentLength**+**L**+$D_{ik}$, where **L** is the latency between the end of the mimicking stimulation sequence and the evoked spike, and $D_{ik}$ is the delay between the mimicked neuron, **i**, and the postsynaptic neuron, **k**.

      - If a weak stimulation falls in the vicinity (i.e. time gap is smaller than **MinGap**) of an existing weak stimulation

         - Merge the two weak stimulations into one strong stimulation, and set the time of the strong stimulation to be the mean time of the two weak stimulations.

         - Update **IsStimulated** (**k**, **SegmentIndex_stimulate**)=1, with **SegmentIndex_stimulate** being the corresponding **SegmentIndex** of the strong stimulation time.

4. Go back to clause 2.

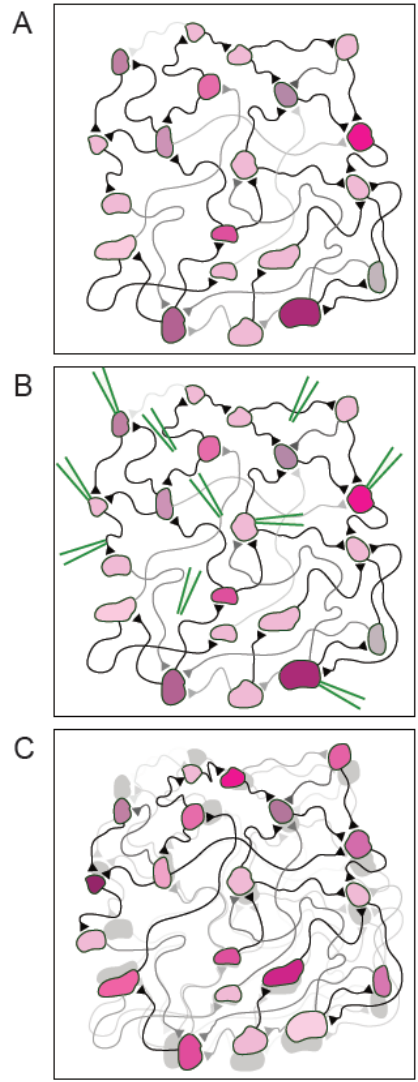

**FIGURE 1 | Illustration of the fundamental experimental difficulty. (A)** An illustration of a neural network. Synaptic strengths and synaptic delays are indicated by the brightness and length of the connections, respectively. The different properties of each neuron are indicated by different colors and shapes. **(B)** The knowledge of the current neuronal and synaptic properties requires an enormous number of measurements carried out by many devices (green), e.g. extracellular and intracellular electrodes, inserted in specific targeted spots in the network. **(C)** The large number of measurements and inserted devices may change the properties of the network, this is schematically exemplified by the difference between the shaded network (identical to the initial network in **(A)**) and the interfered network as a result of the measurements (front colored network).

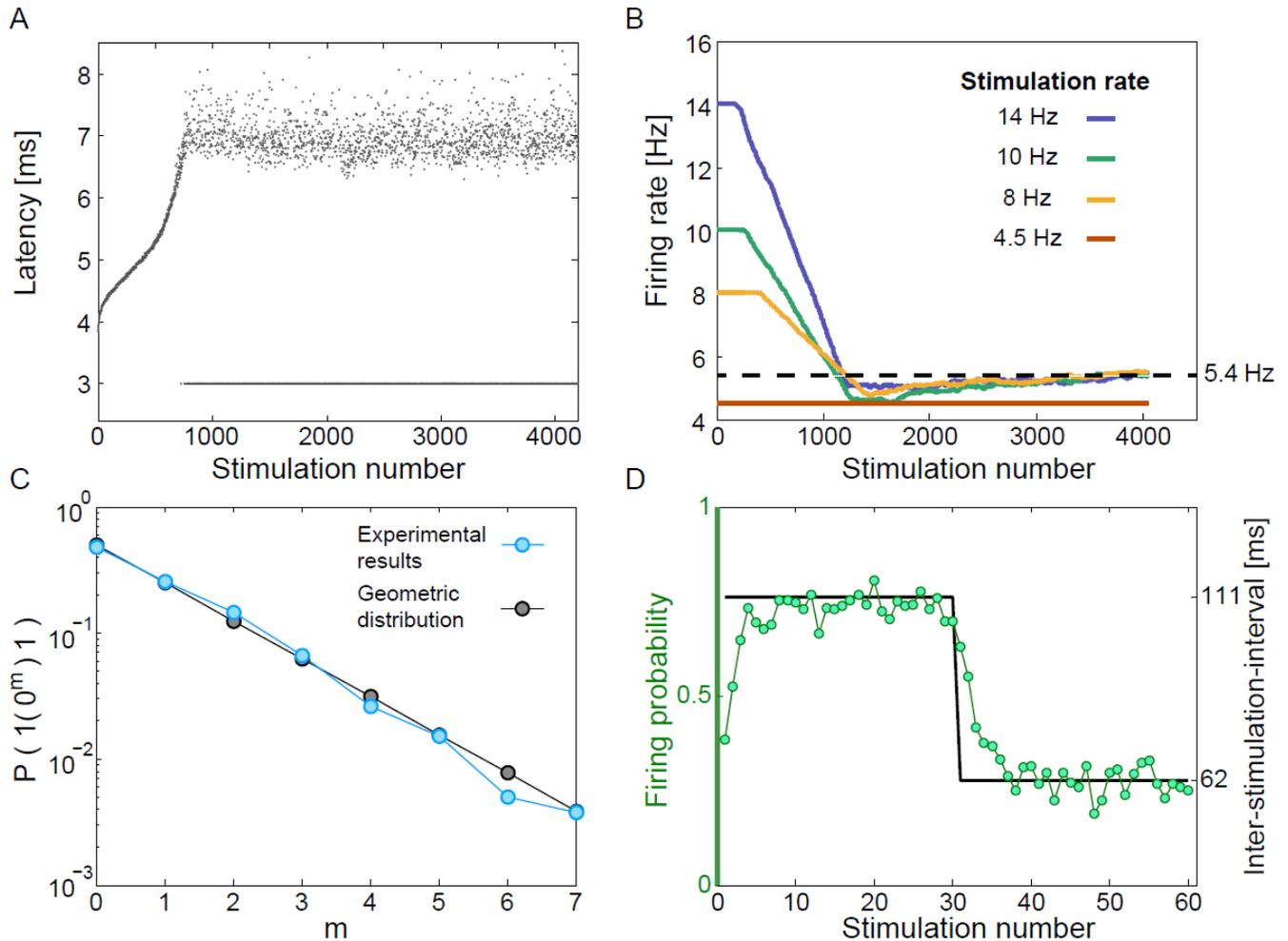

**FIGURE 2 | Neuronal Short-term memory.** **(A)** The neuronal response latency (NRL) of a cultured neuron, stimulated at 10 Hz. Response failures are denoted at NRL=3 ms. **(B)** Firing rates for different stimulation rates (legend), using a sliding window of 1000 stimulations, indicating saturated firing rate (~5.4 Hz, dashed line) independent of the stimulation rate. **(C)** A semi-log plot of the probability for m successive response failures bounded by evoked spikes, as a function of m, for a stimulation rate of 10 Hz (light blue), and for a geometric distribution (P=0.5·0.5^m, black). **(D)** The same neuron was given 233 recurrences of 60 stimulations composed of 30 inter-stimulation intervals of 62.5 ms (16 Hz) and 30 inter-stimulation intervals of ~111 ms (9 Hz) (black). The probability for an evoked spike (green circles) indicates a fast adaptation, short memory, of the neuronal response probability.

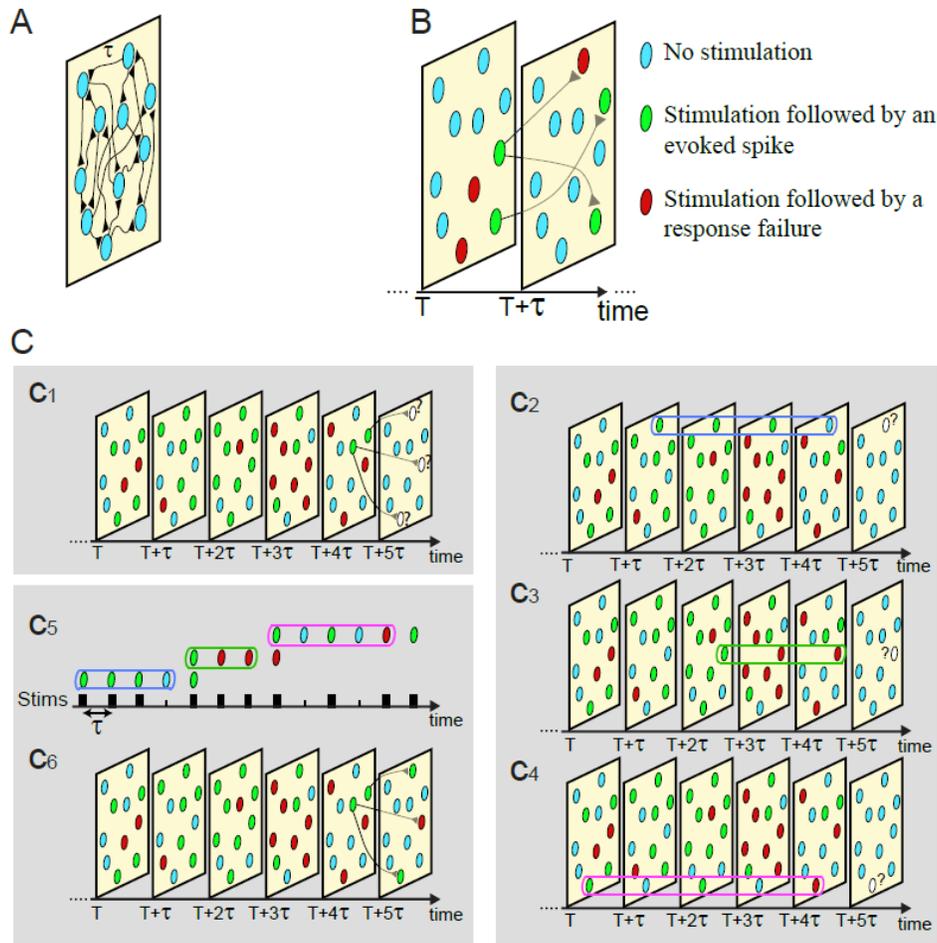

**FIGURE 3 | Illustration of the proposed scheme.** **(A)** An excitatory network where all delays are equal to τ. **(B)** The network dynamics is demonstrated as a set of snapshots for different times, where consecutive snapshots are separated by τ. In each snapshot a neuron can be in one of the following three states: received a stimulation that was followed by an evoked spike (green), received a stimulation that was followed by a response failure (red), or did not receive a stimulation (light blue). **(C)** $C_1$: The state of the neurons in 6 consecutive snapshots of the network, where the current state of the 3 denoted neurons at $T+5\tau$ is unknown. $C_{2-4}$: The responsiveness of each stimulated neuron in the network is determined by its short-term stimulation memory, 3 stimulation in the presented example. $C_{5-6}$: Mimicking the states of these 3 neurons using sequential stimulation of a single neuron, $C_{2-4}$. The responsiveness of the mimicking neuron completely determines the state of the stimulated neurons at layer $T+5\tau$, and the state of the next layer can now be revealed by repeating the described procedure.

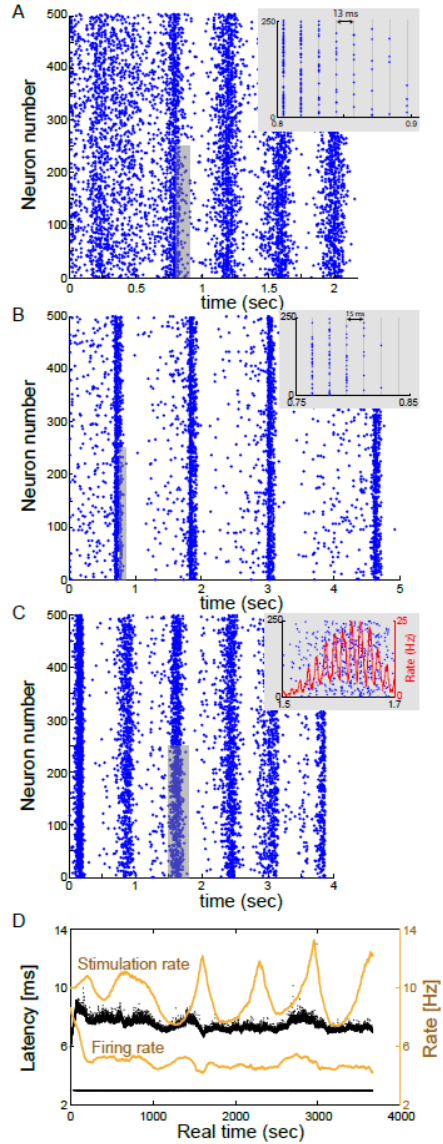

**FIGURE 4 | Utilization of the proposed scheme on excitatory large networks reveals δ and γ oscillations. (A)** Raster plot of a network, where each blue dot indicates an evoked spike, consisting of 500 mimicked neurons where each neuron in the network has randomly selected 2 pre- and 2 post- above-threshold synaptic connections, and all delays are set to 13 ms. Results, produced using a single-neuron experiment *in vitro*, indicate $f_\delta \sim 2.5$ Hz oscillations which coexist with $f_\gamma \sim 75$ Hz oscillations (inset). **(B)** Similar to **(A)** where each neuron has randomly selected 50 pre- and 50 post- below-threshold synaptic connections, and all delays are set to 15 ms. An above-threshold stimulation requires cooperation of at least 4 below-threshold stimulations. Results indicate $f_\delta \sim 0.8$ Hz oscillations which coexist with $f_\gamma \sim 65$ Hz

oscillations (inset). **(C)** A raster plot of a network consisting of 500 neurons where each neuron has randomly selected 2 pre- and 2 post- above-threshold synaptic connections, and delays are randomly selected from the uniform distribution $U(8,12)$ ms. Results indicate $f_\delta\sim1.3$ Hz oscillations which coexist with spontaneous $f_\gamma\sim65$ Hz oscillations, originated from 1/(average($\tau$+latency)) (inset). The rate is calculated from the number of spikes in a sliding window of 20 ms with a resolution of 0.1 ms. **(D)** The NRL of the mimicking neuron in **(A)** (response failures are denoted at NRL=3 ms). The stimulation rate (upper orange curve) and firing rate (lower orange curve) are calculated using a sliding average of 2000 stimulations. The average stimulation rate is much higher than $f_c\sim5$ Hz, indicating that the neuron is in the intermittent phase, which is characterized by large fluctuations of the NRL and response failures which lead to a firing frequency around $f_c\sim5$ Hz.

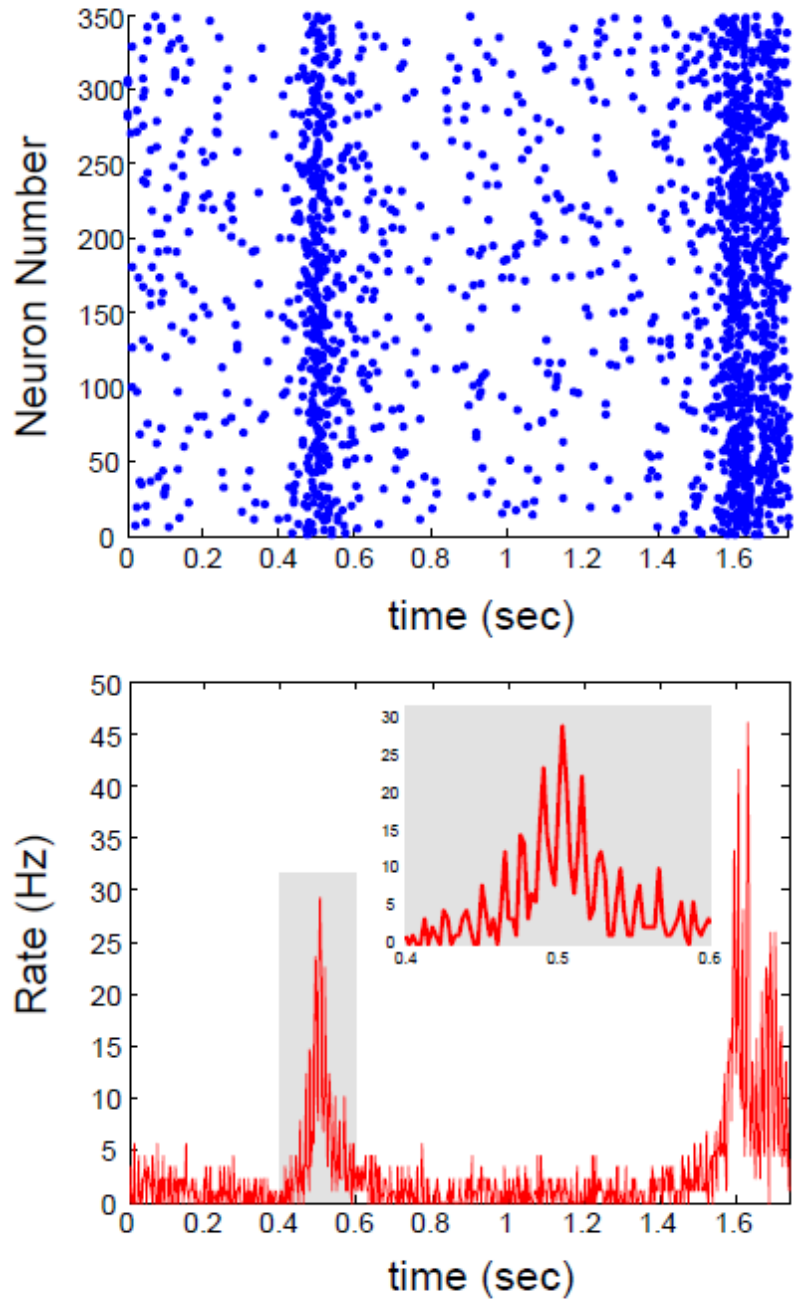

**FIGURE 5 | Sub-threshold connections with continuous delays.** Real-time scheme with continuous delay-times and sub-threshold connections. Top panel: Raster plot of the network activity. Bottom panel: The average firing rate of the neurons comprising the network, as a function of time.

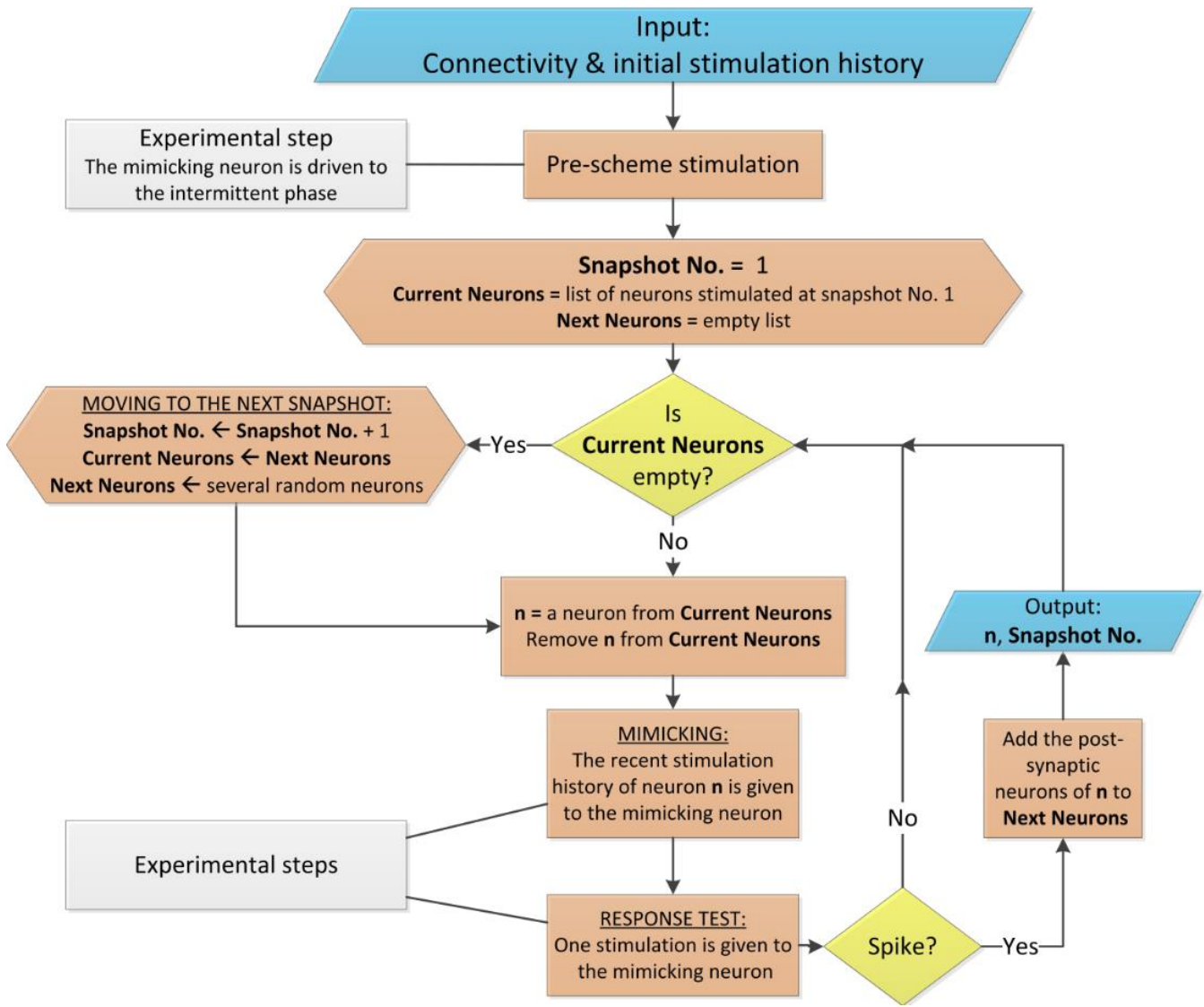

**FIGURE 6 | Flowchart of the proposed scheme.** The flowchart describes the mimicking process for a networks with above-threshold connections and homogenous delays, similar to **Appendix A**. The flowchart uses conventional shapes. Additionally, the colors light blue, yellow, orange and gray stand for data flow, conditional branching, process and experimental comments, respectively. The process is arbitrarily terminated when Snapshot No. is several hundreds (the mimicked time of the network dynamics is several seconds).